\documentclass[manuscript]{emulateapj}
\usepackage{times}
\slugcomment{}

\shorttitle{Field Small Magellanic Cloud stars}
\shortauthors{De Propris et al.}

\begin{document}


\title{A Radial Velocity and Calcium Triplet abundance survey of field Small
Magellanic Cloud giants}


\author{Roberto De Propris\altaffilmark{1}, R. Michael Rich\altaffilmark{2},
Ryan C. Mallery\altaffilmark{2}, Christian D. Howard\altaffilmark{3}}
\email{rdepropris@ctio.noao.edu}


\altaffiltext{1}{Cerro Tololo Inter-American Observatory, La Serena, Chile}
\altaffiltext{2}{Department of Physics and Astronomy, University of California,
Los Angeles, USA}
\altaffiltext{3}{SOFIA Science Centre, Moffett Field, California, USA}


\begin{abstract}

We present the results of a pilot wide-field radial velocity and metal
abundance survey of red giants in ten fields in the Small Magellanic Cloud 
(SMC). The targets lie at projected distances of 0.9 and 1.9 kpc from the 
SMC centre ($m-M=18.79$) to the North, East, South and West. Two more fields 
are to the East at distances of 3.9 and 5.1 kpc. In this last field we find 
only a few to no SMC giants, suggesting that the edge of the SMC in this direction 
lies approximately at 6 kpc from its centre. In all eastern fields we observe a 
double peak in the radial velocities of stars, with a component at the classical 
SMC  recession velocity of $\sim 160$ km s$^{-1}$ and a high velocity component 
at about 200 km s$^{-1}$, similar to observations in H{\small I}. In the most 
distant field (3.9 kpc) the low velocity component is at 106 km s$^{-1}$. The 
metal abundance distribution in all fields is broad and centred at about [Fe/H] 
$\sim -1.25$, reaching to solar and possibly slightly supersolar values and down 
to [Fe/H] of about $-2.5$. In the two innermost (0.9 kpc) Northern and Southern 
fields we observe a secondary peak at metallicities of about $\sim -0.6$. This 
may be evidence of a second episode of star formation in the centre, possibly 
triggered by the interactions that created the Stream and Bridge.
 
\end{abstract}





\section{Introduction}

The Small Magellanic Cloud (hereafter SMC) is, together with the Large Magellanic
Cloud (LMC) the nearest dwarf irregular galaxy to our own, and provides an invaluable 
laboratory to study star formation and chemical evolution in low mass galaxies.
There is recent evidence that the LMC and SMC are on their first pass around the
Milky Way and that the SMC may not be bound to the LMC \citep{kallivayalil06a,
kallivayalil06b,besla07}. The SMC may be a rare example of a comparatively isolated
dwarf galaxy and possibly even a surviving fragment from the era of reionization. 
However, the SMC has also been interacting with the LMC during the past few Gyrs 
and these interactions have modulated the recent star formation history of both
galaxies (e.g., \citealt{bekki05,bekki09} and references therein). 

The SMC is best modelled as an old dwarf spheroidal galaxy possessing a gaseous disk 
\citep{bekki09} that has been distorted by star formation and tidal stresses, giving 
the galaxy its present irregular appearance (e.g., \citealt{harris04,cioni06}). The 
distribution and chemical abundances of field stars in the SMC thus provide clues 
to its star formation history. Open questions include: whether there is an `edge' 
to the SMC, the metallicity distribution of its field stars, the presence of a metal 
abundance gradient and whether a metal-poor halo exists around the SMC or other dwarf 
galaxies as it does around the Milky Way and other giants.

Stars belonging to the SMC have been found along the Magellanic Bridge; an
old and intermediate age population out to 5$^{\circ}$ but only a young population
at $\sim 6.5^{\circ}$ \citep{harris07}. \cite{noel07} explored three fields to 
the South of the SMC identified SMC stellar sequences belonging to the intermediate 
age population out to 6.5 kpc from the SMC centre. In other galaxies, \cite{munoz06} 
observed LMC stars as far as 23$^{\circ}$ from its centre. Extended stellar envelopes 
are also detected around other dwarfs (e.g., \citealt{minniti96,vansevicius04,hidalgo09},
but at least in some cases, these are actually tidal in origin \citep{munoz06}. Although 
stars are proved to exist at large projected distances in many nearby dwarfs, these 
objects may not represent a classical metal-poor halo as is encountered in the Milky Way 
or M31. For instance, in the LMC stars studied by \cite{munoz06}, the metallicity distribution 
is broad and centred around [Fe/H] $\sim -1$, with a large range of ages \citep{gallart04},
unlike the largely old and metal-poor stars that are believed to populate the outer halos
of giant galaxies. The SMC itself appears to have formed stars quickly at early epochs 
reaching a metallicity of $\sim -1$ and to have then suffered a series of star formation 
episodes over the past 3 Gyrs, after a period of quiescence, which have produced younger 
stellar populations and more metal rich stars \citep{harris04}. 

In the innermost regions of the SMC, \cite{carrera08} found an average metallicity
of [Fe/H] $\sim -1$, in agreement with previous studies, but also claimed to have
detected a metal abundance gradient (richer inward), arguing that this is related to 
an age gradient, with younger (and more metal rich) stars towards the SMC centre. 
While this agrees with the earlier work of \cite{piatti07a,piatti07b}, the study of 
SMC clusters and field giants (in the proximity of clusters) by \cite{parisi08,parisi10}, 
as well as the work by \cite{cioni09} on the C/M ratio of AGB stars in the SMC, do not 
support the existence of the metal abundance gradient claimed by \cite{carrera08}.

In this {\it Letter} we report on a pilot program for an extensive radial velocity and 
Calcium Triplet survey of the SMC, based on data collected during a similar survey
of the Galactic Bulge. The observations and data reductions are described in the
next section, while we present the main results and our discussion in the following
sections. We adopt the most recent distance modulus of 18.79 for the SMC \citep{
sz09}.

\section{Observations}

Data for this project were obtained as part of the Bulge Radial Velocity Assay
(BRAVA -- \citealt{rich07,howard08,howard09}). The BRAVA survey was 
allocated a series of runs between August 2008 and August 2009 to carry
out radial velocity measurements of $\sim 10000$ M giants in the Bulge of
the Milky Way. The allocated times were somewhat sub-optimal for bulge
observations and left about 1/3 of each night free, after the Bulge sank to
high airmass at about 02:30. We decided to dedicate this remaining time to
an exploratory survey to study the kinematics and chemical abundances of
K giants in the SMC.

Observations were taken at the V. M. Blanco 4m telescope on Cerro Tololo,
Chile, using the Hydra multi-fiber spectrograph. We used the KPGLD grating
(790 l/mm, blazed at 8500 \AA), with the 200 $\mu$m slit mask to achieve a 
resolution of 4200, covering a spectral range of 1800 \AA\ centred on 7900
\AA, including the three Calcium Triplet (CaT) lines at 8498, 8542 and 8662
\AA. Further details on the survey may be found in the BRAVA paper by
\cite{howard08}. Exposure times for each target were 3 $\times 1200$
seconds.

Targets were selected from the 2MASS database \citep{skrutskie06} as
luminous red giants at the distance of the SMC, with $13 < K < 14$ and 
$0.5 < J-K < 1.5$. The selection range and method are analogous to those
used for the Galactic Bulge BRAVA survey. Figure 1 shows our targets
overplotted over the Padova isochrones \citep{marigo08} shifted to the
assumed SMC distance.

\begin{figure}
\plotone{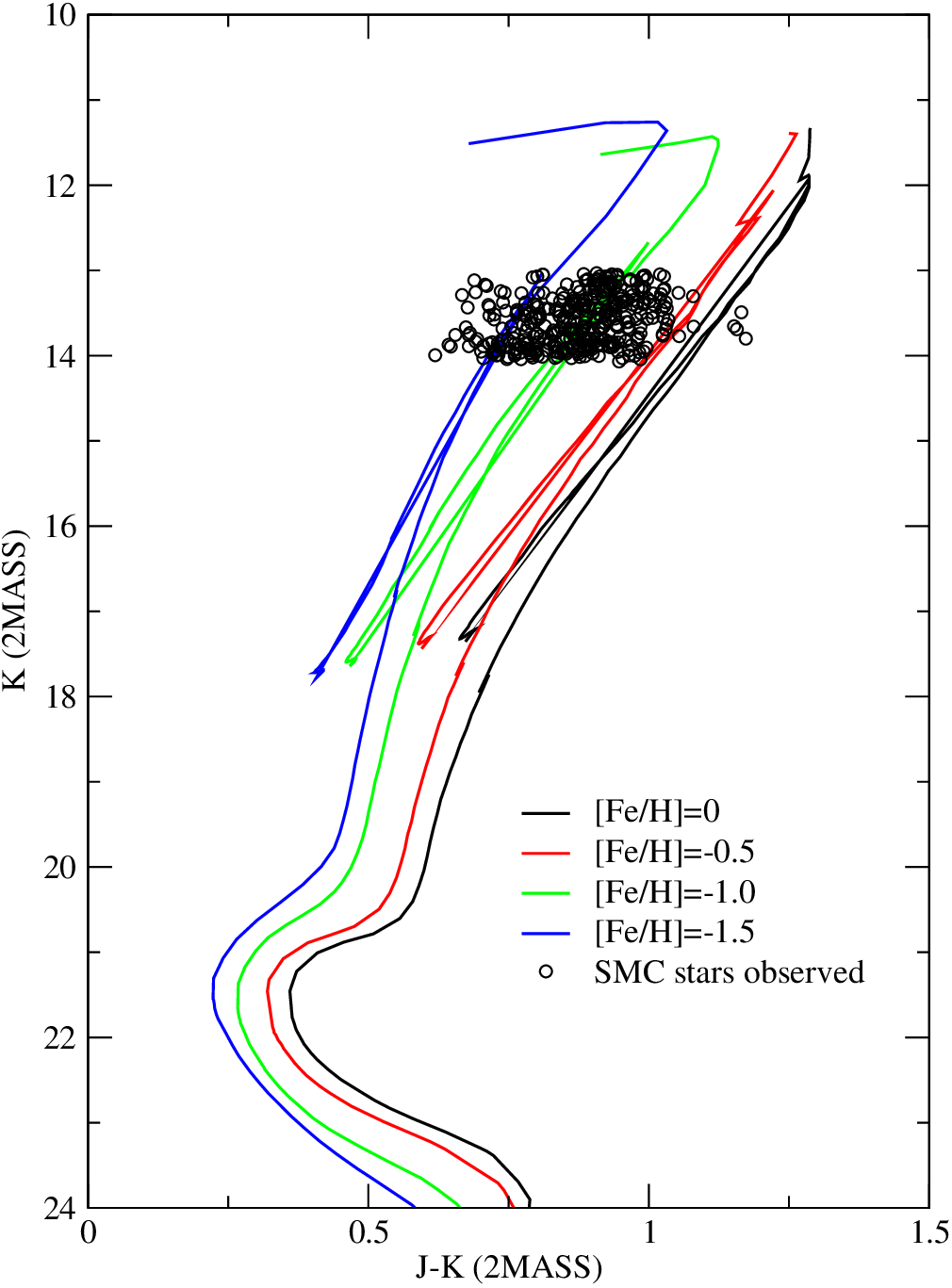}
\caption{The target stars shifted on to Padova isochrones of
age 10 Gyr with metallicities as indicated in the legend, 
assuming a SMC distance modulus of 18.79 mag.}
\end{figure}

Table 1 shows, in column order, the identification of the fields, their
positions of the fields (equinox 2000), their distances from the SMC centre 
(in kpc, assuming the above distance modulus), the number of stars we attribute 
to the SMC (see below for the definition of the velocity range we considered), 
the number of stars with successful redshifts we obtain and the total number of 
stars surveyed. Figure 2 plots the positions of the fields on the sky with respect
to the SMC (using 2MASS data). We have one field at each of the four cardinal points, 
at projected distances of 0.9 and 1.9 kpc, and two fields to the East of the SMC, 
at distances of 3.9 and 5.1 kpc.

\begin{deluxetable*}{lcccccc}
\tablecaption{Positions of Observed Fields}
\tablewidth{0pt}
\tablehead{
\colhead{Field Name} & \colhead{RA (2000)} & \colhead{Dec (2000)} & 
\colhead{R$_{SMC}$ (kpc)} & \colhead{N$_{SMC}$} & \colhead{N$_{spectra}$} & 
\colhead{N$_{targets}$}
}
\startdata
5001 & 01:05:42.2 & $-72$:50:31.4 & 0.96 & 81 & 82 & 105 \\
5002 & 01:18:33.8 & $-72$:48:41.8 & 1.91 & 82 & 94 & 101 \\
5003 & 01:45:44.4 & $-72$:50:30.4 & 3.90 & 43 & 63 & 81 \\
5004 & 02:01:31.5 & $-72$:47:17.8 & 5.06 & 5 & 44 & 62 \\
5005 & 00:40:47.7 & $-72$:53:36.6 & 0.88 & 57 & 74 & 98 \\
5006 & 00:26:30.1 & $-72$:51:44.2 & 1.93 & 58 & 61 & 108 \\
5009 & 00:53:01.5 & $-71$:53:26.8 & 0.94 & 75 & 86 & 117\\
5010 & 00:52:40.9 & $-70$:53:43.4 & 1.94 & 36 & 57 & 91 \\
5013 & 00:52:34.2 & $-73$:43:53.4 & 0.90 & 78 & 82 & 112 \\
5014 & 00:52:57.9 & $-74$:44:43.4 & 1.91 & 36 & 40 & 100 \\
\enddata
\end{deluxetable*}

\begin{figure*}
\vspace{2cm}
\plotone{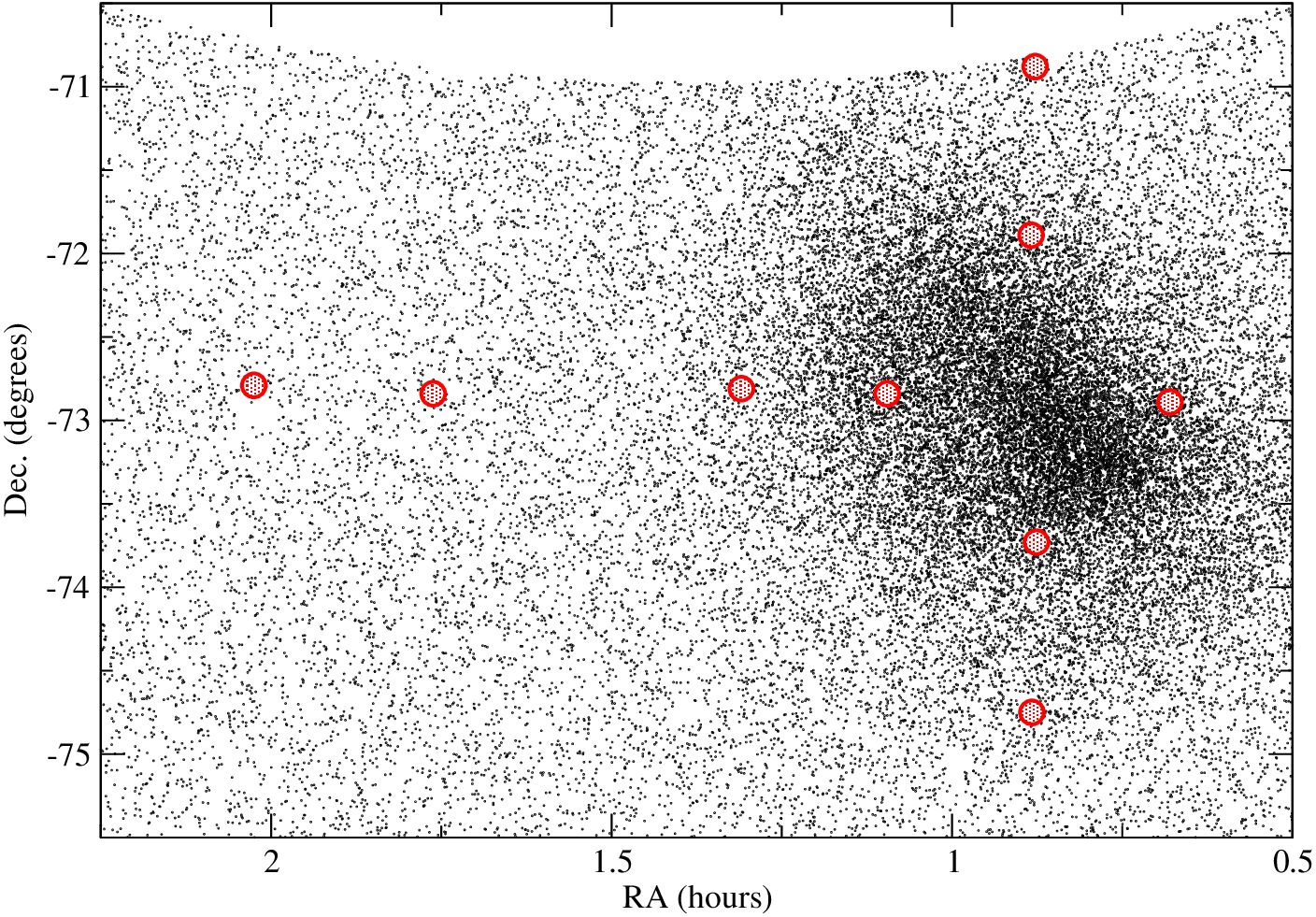}
\caption{Map of the SMC from 2MASS data, using stars with $13 < K < 14$ and
$0.5 < J-K < 1.5$. We overplot circles representing the fields observed and
whose positions are reported in Table 1.}
\vspace{1cm}
\end{figure*}

We reduced the data by first removing the overscan, trimming and then
carrying out bias subtraction. At this point we rejected cosmic rays with
the Laplacian Edge algorithm \citep{vandokkum01}. We then extracted
the spectra using the specialized IRAF package {\tt dohydra} and finally
combined all the individual spectra for each star using the {\tt scombine}
task.

We derived radial velocities by cross-correlating our data with spectral
templates \citep{tonry79} focusing on the region containing the CaT 
lines. We then measured the CaT index \citep{cenarro01} for our stars
using the {\sl indexf} program \citep{cardiel07} and converted our 
measurements to [Fe/H] using the fitting functions tabulated in \cite{
cenarro02}, with $T_{eff}$ and $\log g$ from the Padova isochrones
\citep{marigo08}. Our stars are actually comparatively faint, with $I$ 
of 16 to 17, plus we often observed after moonrise, so our errors are 
relatively large: $\pm 10$ km s$^{-1}$ in radial velocity and $\pm 0.1$
dex in [Fe/H].

\section{Kinematics}

In Figure 3 we show radial velocity histograms for all fields (as marked
in the figure legend) and a comparison with expectations from the Besancon
model of the Galaxy \citep{robin03}, assuming the same selection criteria
as our SMC fields and over the $40'$ field of view sampled by Hydra.

\begin{figure}
\epsscale{0.8}
\plotone{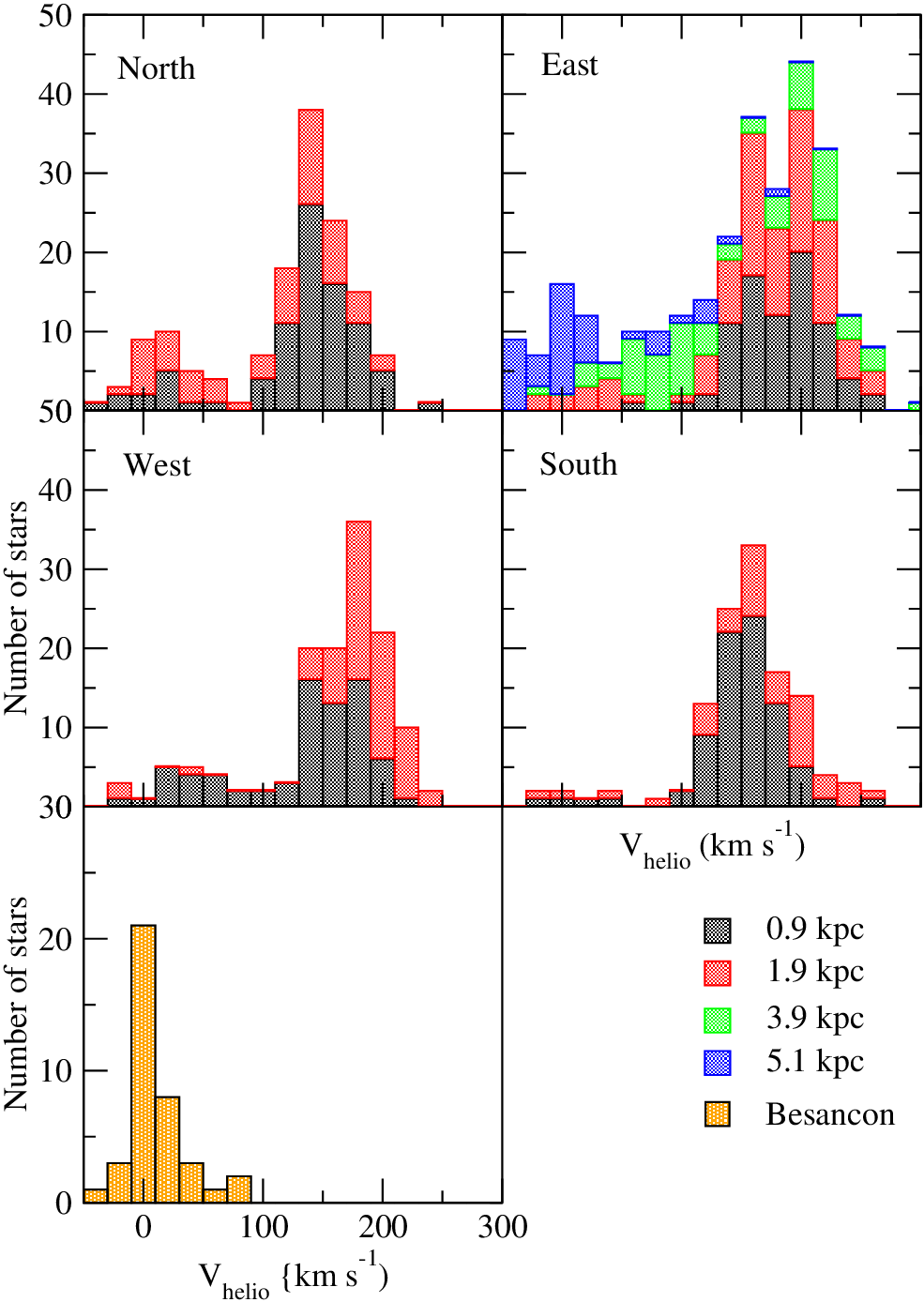}
\caption{Histograms of radial velocities for observed fields. The
data are plotted as stacked charts with a different color for each
distance surveyed and the colors are identified in the figure legend. 
We plot the North, South, East and West fields in different panels as
identified in the figure. The bottom panel shows the predictions of a Besancon
model of the Galaxy for the same selection criteria as our targets
and over a single Hydra field.}
\end{figure}

We assume that all stars with heliocentric radial velocity greater than 100
km s$^{-1}$ can be attributed to the SMC, as implied by the distribution of
velocities in the Besancon model (Figure 3). In the Northern 0.9 kpc field, 
the stars follow a Gaussian distribution with $<v>=152 \pm 26$ km s$^{-1}$, 
while the South 0.9 kpc field has $<v>=160 \pm 39$ km s$^{-1}$. These values 
are broadly consistent with previous measurements as tabulated by \cite{harris04}.
However our East 0.9 kpc field shows a bimodal distribution. The KMM algorithm 
\citep{ashman94} returns two peaks at 162 and 205 km s$^{-1}$ with 72\% significance. 
In the Western 0.9 kpc field there also appears to be a bimodal distribution, with 
peaks at 143 and 180 km s$^{-1}$ but at much lower significance. The lower
velocity peak is consistent with the conventional SMC recession velocity,
while the other peak may correspond to a high velocity component, detected in
H{\small I} observations \citep{mathewson84,stanimirovic04}.

For the 1.9 kpc fields, the Northern field has $<v>=147 \pm 26$ km s$^{-1}$,
while for the Southern field the distribution appears to be bimodal with
peaks at 156 and 206 km s$^{-1}$ with a 75\% significance according to 
the KMM algorithm. As for the 0.9 kpc field, the Eastern field also appears
to show a bimodal velocity distribution, with peaks at 162 and 211 km s$^{-1}$
at a 58\% significance level. The Western field is instead consistent with
a single component with $<v>=189 \pm 23$ km s$^{-1}$. In all these cases we
recover the `typical' SMC velocity dispersion of about 25 km s$^{-1}$.

We also observed two fields to the East, at a distance of 3.9 and 5.1
kpc from the SMC centre. The SMC is clearly present in the 3.9 kpc
field. This has a pronounced bimodal velocity distribution, with peaks
at 106 and 211 km s$^{-1}$. While the high velocity component is at
approximately the same radial velocity in all eastern fields, and in
the Southern field as well, the velocity separation with the lower
velocity component appears to increase significantly outwards from
the SMC centre. 

On the other hand, the SMC contribution is very small to non-existent
in the most distant field to the East. This suggests that the edge
of the SMC in this direction is close to this field. To estimate this,
we first calculated the number of SMC stars in each of our Eastern
fields based on the derived completeness and fraction of stars we
attribute to the SMC in Table 1. We then integrated a 10 Gyr old, 
[Fe/H]=$-1.0$ luminosity function from \cite{marigo08} and derived the 
surface brightness of SMC stars in our fields, after correcting for 
incompleteness and sampling. We then fit this to a Hernquist profile 
and extrapolated to a surface brightness of 26 mag arcsec$^{-2}$ in $K$, 
which we take, arbitrarily, as the SMC `limit'. This exercise returns a 
distance of 5.8 kpc, unlike what observed by \cite{harris04} along the 
Magellanic Bridge, and the detection of SMC stars out to about 6.5 kpc  
to the South of the SMC by \cite{noel07}. These latter stars may of course have been 
formed and/or tidally displaced from the SMC by the repeated interactions 
with the LMC.

\section{Metallicities}

Figure 4 shows the distribution of metal abundances, from the Calcium Triplet,
in the 8 inner fields. At a projected distance of 0.9 kpc from the SMC
centre, both the East and West fields show broad metallicity distributions
centered at [Fe/H]=$=1.27 \pm 0.05$ with a dispersion of $0.48 \pm 0.03$
and $-1.47 \pm 0.05$ with a dispersion of $0.99 \pm 0.05$ respectively,
consistent with previous studies \citep{piatti07a,piatti07b,carrera08,
parisi08,parisi10}. The metallicity distribution for the 0.9 kpc
field to the North instead appears to exhibit a peak at about [Fe/H]
$\sim -1.3$ and another at [Fe/H] $\sim -0.6$, while in the South
the metallicity distribution is also broad and extending to solar
or slightly supersolar metal abundances. The metallicity distributions
are also similar in the four 1.9 kpc fields, but the higher metallicity
stars in the North and South fields are no longer present. All these
fields have [Fe/H] $\sim -1.35 \pm 0.10$ with a dispersion of about
$0.65 \pm 0.08$.

\begin{figure}
\epsscale{0.7}
\plotone{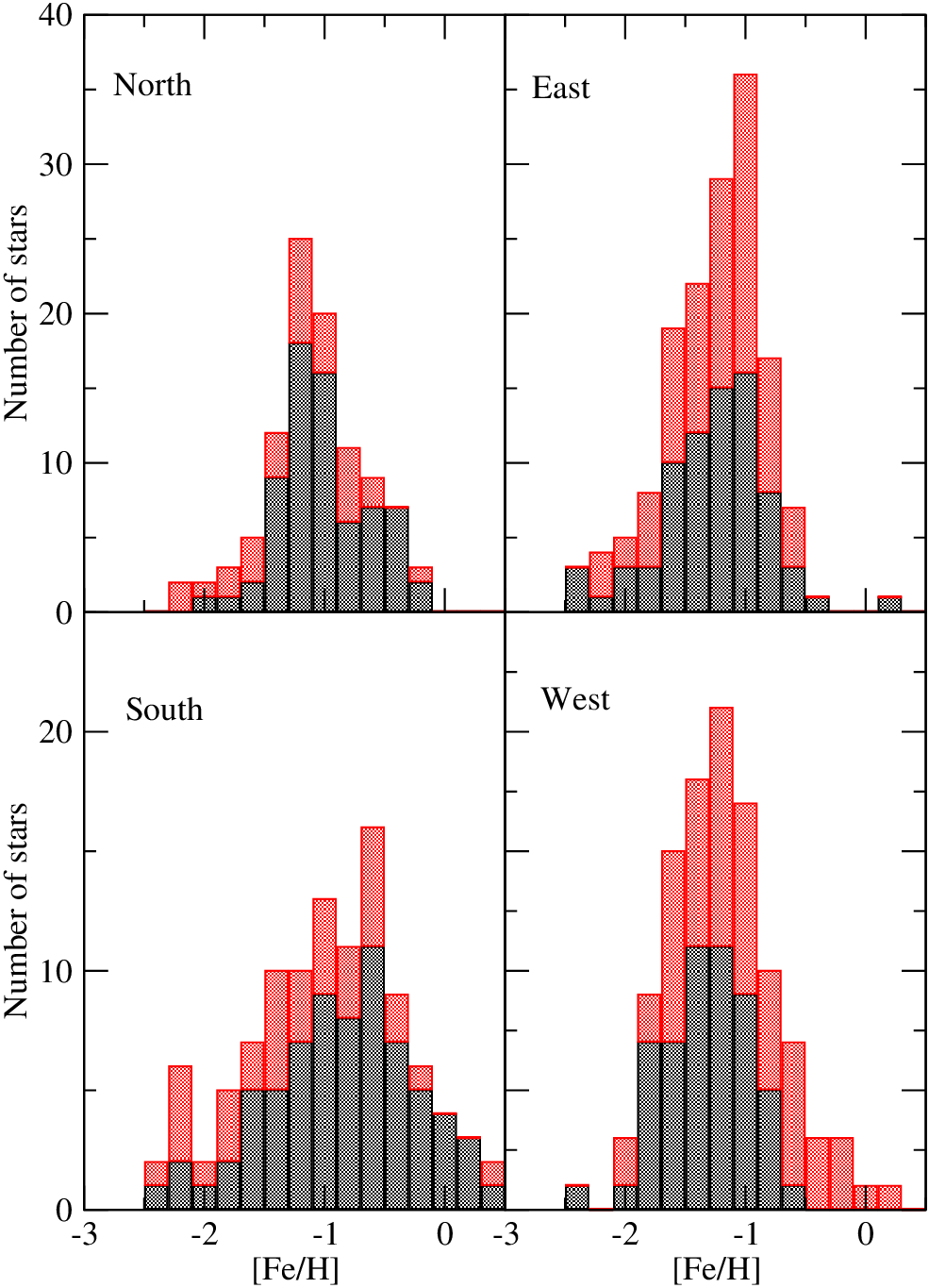}
\caption{The metallicity distribution (measured via the CaT) of stars in
the fields at 0.9kpc and 1.9kpc from the SMC centre. We plot our fields as
in Figure 3. Identifications are as in the legend in Figure 3.}
\end{figure}

While we do not see a radial abundance trend, as claimed by \cite{carrera08}
the disappearance of the higher metallicity stars in the outer Northern and
Southern fields, with respect to the inner fields, may explain the discrepancy
between the claim of \cite{carrera08} for a metal abundance gradients and
the results of \cite{parisi08,parisi10} and \cite{cioni09}. While the majority
of the SMC stars follow a broad metallicity distribution with no radial
trend, some of the inner fields contain a more metal rich, and presumably
younger, population. This would mimic a metal abundance gradient if the
bimodality of the distribution is not taken into account.

\section{Discussion}

The data we present here show that, while the SMC is detected to large
distances (about 6 kpc) along the Magellanic Bridge \citep{harris04} and
to the South \citep{noel07}, we appear to have approached the edge of
the SMC in our easternmost fields. We estimate that the SMC `edge' in
this direction lies at about 6 kpc from its centre. The shape of the 
SMC has however been tidally distorted by interactions with the LMC and 
it has been elongated along the N-S direction \citep{kunkel00}. Exploration 
of the radial and azimuthal behavior at larger distances will be one of 
the outcomes expected from a wider-field spectroscopic survey.

The kinematics of stars in our fields is complex. There is evidence for
the presence of two components in some fields, particularly to the East
and the South, i.e. the regions most affected by interactions with the LMC
and the Magellanic Bridge: a low velocity one around 160 km s$^{-1}$
and a high velocity component at about 210 km s$^{-1}$. Although the
evidence is weak, while the high velocity component is at the same
position in our 3.9kpc field to the East, the low velocity component
appears to have lower velocity. This is reminiscent of the claims
for multiple peaks in the H{\small I} velocity distribution \citep{
mathewson84,stanimirovic04}. Similar bifurcations are also observed in 
tidal streams. The H{\small I} features are often attributed to multiple and
overlapping gas shells, but their presence in the stellar distributions,
especially in the zones to the East and South closer to the Magellanic
Bridge, may favor multiple components models such as those of
\cite{mathewson84}. 

An intriguing possibility is that we are detecting stars from the LMC in
the SMC Eastern fields: \cite{munoz06} find the presence of LMC stars as far 
23$^{\circ}$ from the LMC centre (which of course lies to the East of the
SMC), while \cite{bekki08} has argued for the existence of a common halo
encompassing the LMC and SMC. Based on the recent study of LMC kinematics
by \cite{vandermarel02}, we would expect LMC stars to lie at the position
of the second velocity peak we observe in the Eastern fields. This seems
somewhat less likely because we find that the secondary peaks in our data
contain about the same number of stars as the primary velocity peaks,
and therefore appear more likely to be associated with the SMC velocity
structure than the LMC. Since our easternmost fields are closer to the
LMC than the 0.9 kpc eastern field, we would expect the LMC contribution
(if it causes the secondary peak) to increase `outward' from the SMC,
unlike the observations. It is of course very likely that some LMC
stars are actually superposed over the SMC, but they can probably be
securely separated out only by chemical tagging.

One striking feature we observe in our data is a broad metallicity distribution
centered on [Fe/H] $\sim -1.2$, extending from $-2.5$ to solar or
even slightly supersolar values. This is very similar to what observed in 
the LMC by \cite{munoz06} and in Sagittarius by \cite{monaco05} and may 
suggest a very similar chemical evolution pattern in most dwarf galaxies. 
In fact the abundance distribution we observe in the SMC is also very similar 
to that measured for the M31 ``giant stream'' population, with a peak near 
[Fe/H] $\sim -1$ and tails to high and low  metallicities \citep{koch08}. 
The ingestion of massive galaxies such as the SMC has been invoked to explain 
the wide metallicity distribution and the presence of metal-rich stars in the 
M31 halo (see e.g., \citealt{koch08}). In the case of the SMC there is the 
question of how a galaxy massive enough to host such metal-rich stars could 
have been accreted by the SMC without more significant disruption of the SMC 
(but see \citealt{tsujimoto09}). 

The broad metallicity distribution may instead imply the presence of multiple 
stellar generations. It is known that the SMC has undergone recent star formation,
possibly induced by encounters with the LMC, after a long period of quiescence
\citep{harris04}. \cite{carrera08} claim that there is a metal abundance 
gradient in the SMC and suggest that this is due to the presence of younger
stars in the centre of this galaxy. We find that the main population of the
SMC does not exhibit a metal abundance gradient \citep{parisi08,parisi10}, 
but that in the inner fields to the North and South there is a contribution 
from more metal-rich stars, with peak metallicity around [Fe/H] $\sim -0.6$. 
\cite{noel07} find evidence for an intermediate age population in their fields 
to the South out to 6.5 kpc, while a younger stellar population is detected 
by \cite{harris04} along the Magellanic Bridge. The presence of more metal-rich 
stars forming a separate peak in the inner fields resembles the picture of 
\cite{carrera08} where a recent burst of star formation has led to self-enrichment 
in the inner regions. The approximate North-South trend is roughly in the
directions of the Bridge and Stream features and it is tempting to speculate
that the interactions that created these gaseous features are also responsible
for the star formation episodes.

A wider and larger spectroscopic survey will allow us to clarify the structure
and kinematics of the SMC, explore the existence of metallicity gradients,
search for a metal poor halo and detect the presence of streams.

%% In a manner similar to \objectname authors can provide links to dataset
%% hosted at participating data centers via the \dataset{} command.  The
%% second curly bracket argument is printed in the text while the first
%% parentheses argument serves as the valid data set identifier.  Large
%% lists of data set are best provided in a table (see Table 3 for an example).
%% Valid data set identifiers should be obtained from the data center that
%% is currently hosting the data.
%%
%% Note that AASTeX interprets everything between the curly braces in the 
%% macro as regular text, so any special characters, e.g. "#" or "_," must be 
%% preceded by a backslash. Otherwise, you will get a LaTeX error when you 
%% compile your manuscript.  Special characters do not 
%% need to be escaped in the optional, square-bracket argument.

%% In this section, we use  the \subsection command to set off
%% a subsection.  \footnote is used to insert a footnote to the text.


%% This section also includes several of the displayed math environments
%% mentioned in the Author Guide.

\acknowledgments

This research has made use of the NASA/ IPAC Infrared Science Archive, which is operated by the Jet Propulsion Laboratory, California Institute of Technology, under contract with the National Aeronautics and Space Administration. This publication makes use of data products from the Two Micron All Sky Survey, which is a joint project of the University of Massachusetts and the Infrared Processing and Analysis Center/California Institute of Technology, funded by the National Aeronautics and Space Administration and the National Science Foundation. We wish to thank the anonymous referee for a very helpful report that
has substantially helped us to improve the paper.


%% in parentheses, after the keyword, but they will not be verified.

{\it Facilities:} \facility{CTIO (Hydra)}.

\clearpage

\end{document}